# Solid Lubrication with MoS$_2$: A Review

**Mohammad R. Vazirisereshk [1], Ashlie Martini [1], David A. Strubbe [2] and Mehmet Z. Baykara [1,\*]**

[1] Department of Mechanical Engineering, University of California Merced, Merced, CA 95343, USA; mvazirisereshk@ucmerced.edu, amartini@ucmerced.edu
[2] Department of Physics, University of California Merced, Merced, CA 95343, USA; dstrubbe@ucmerced.edu
\* Correspondence: mehmet.baykara@ucmerced.edu

**Abstract:** Molybdenum disulfide (MoS$_2$) is one of the most broadly utilized solid lubricants with a wide range of applications, including but not limited to those in the aerospace/space industry. Here we present a focused review of solid lubrication with MoS$_2$ by highlighting its structure, synthesis, applications and the fundamental mechanisms underlying its lubricative properties, together with a discussion of their environmental and temperature dependence. An effort is made to cover the main theoretical and experimental studies that constitute milestones in our scientific understanding. The review also includes an extensive overview of the structure and tribological properties of doped MoS$_2$, followed by a discussion of potential future research directions.

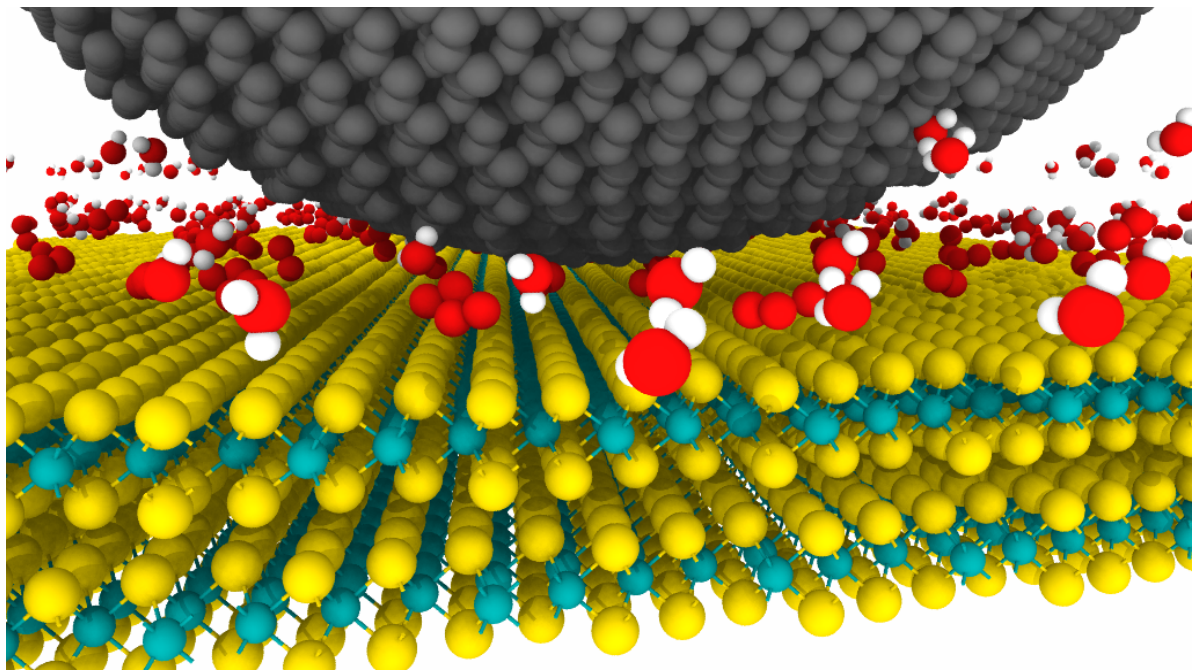

## 1. Introduction

With their notable influence on an overwhelming number of natural and technical processes, friction, wear and lubrication are subjects that are worthy of intense scientific investigation. Despite their importance, studying individual aspects of these phenomena (which are influenced by a multitude of factors including the structural, mechanical and chemical properties of the involved surfaces) in an isolated fashion has been a challenging endeavor. Nevertheless, with improved computational and experimental techniques, *tribology* (a term coined in the 1960s that describes the study of friction, wear and lubrication) is now an active field of inquiry, with research groups around the world working on understanding and utilizing its various aspects [1,2].

Considering that a few percent of the gross domestic product of a developed nation is wasted on overcoming the negative impacts of friction and wear (in the form of direct loss of useful energy as well as replacement costs associated with machine components damaged by wear) [3], a major goal of tribology research involves the design and application of methods to minimize friction and wear at interfaces in relative motion. Toward this goal, significant effort is spent on (i) understanding the properties and improving the usefulness of existing lubricants and (ii) coming up with new ways of lubricating interfaces in a more effective fashion (e.g. by decreasing the coefficient of friction, reducing wear rate and improving the lifetime of components).

Liquid lubricants are typically what comes to mind when one thinks of lubrication in an industrial setting [4]. It is indeed true that oils and greases are conventionally employed in many mechanical systems for lubrication purposes, mostly due to the robust reduction in friction and wear rate that they provide, the ease with which they can be replenished and the straightforward manner with which they can be applied to a variety of surfaces. However, certain operating conditions strictly require the use of *solid lubricants* instead [5]. The most common examples involve aerospace/space applications, where low temperatures preclude the use of liquid lubricants which simply become too viscous for effective lubrication or may even freeze [6,7]. In fact, most of the technical progress involving solid lubricants was achieved in the second half of the 20$^{th}$ century, motivated by the advent of the space age. Other examples of application for solid lubricants involve dry machining operations, where the use of solid lubricants on machine tools (mainly for wear reduction purposes) may significantly cut down on the cost of using large amounts of liquid lubricants, and also the food processing and textile industries, where contamination by liquid lubricants may pose health hazards [8]. In general, solid lubricants are suitable for many applications that operate in conditions too extreme for liquid lubrication.

Among solid lubricants currently in use, molybdenum disulfide ($MoS_2$) [9] holds special importance. Being a *lamellar* solid material that consists of individual atomically-thin planes that can easily slide against each other, the use of $MoS_2$ as a solid lubricant in modern technology dates back to the previous century [10]. Like graphite (another lamellar material), $MoS_2$ can be used as a dry lubricant by itself, as an additive in oils or greases, or as an individual component of a composite coating [2]. Unlike graphite, $MoS_2$ does not require humid environments to perform well and it has indeed been shown that its lubricative properties improve drastically in oxygen-deficient environments [11]. Combined with its ability to operate reliably in a wide range of temperatures (from the cryogenic regime to several hundred degrees Celsius), the ability to function effectively in vacuum makes $MoS_2$ a particularly attractive lubricant for aerospace/space applications. Consequently, a great body of scientific work exists on the synthesis, properties and applications of $MoS_2$ as a solid lubricant and reviews as well as books on the topic have been published as early as the 1960s [8,9,12,13]. However, there is still active research on understanding and improving the lubricative properties of $MoS_2$, with some of the recent focus areas being on its incorporation into (nano-)composite coatings and improving its properties via controlled doping [14,15].

This review (which by no means is intended to be comprehensive) provides a focused overview of the tribology of $MoS_2$ in the form of a snapshot of our current physical understanding of this exciting solid lubricant, with the hope that it could potentially guide future fundamental work. In alignment with this mindset, we focus specifically on $MoS_2$ by itself as a solid lubricant and do not cover the rather extensive body of work related to its use as an additive in oils and greases or its utilization in (nano-)composite coatings. Specifically, Section 2 describes the structure and synthesis of $MoS_2$, while Section 3 includes an overview and specific examples of its applications as a solid lubricant. Section 4 details the fundamental mechanisms of low friction and wear associated with the material, followed by Sections 5 and 6 which discuss the environmental and temperature dependence of its lubricative properties, respectively. Section 7 deals with doped $MoS_2$ and, finally, Section 8 concludes the review by providing a brief summary along with highlights of some of the emerging research directions for $MoS_2$ as a solid lubricant.

## 2. Structure and Synthesis

$MoS_2$ belongs to the family of layered two-dimensional transitional metal dichalcogenides (TMDs). Like graphite and hexagonal boron nitride, its crystal structure consists of covalently bonded sheets, which form stacks that are held together only by weak Van der Waals interactions [16]. It occurs naturally as the mineral molybdenite, an important Mo ore. Due to the strong bonding in-plane and weak bonding out-of-plane, mechanical and other properties are highly anisotropic [17], and the stacks can be easily sheared. Single-layer or few-layer (i.e. 2D) $MoS_2$ (analogous to graphene) can be produced and studied individually [18].

$MoS_2$ has several different possible structures (polytypes), depending on the bonding within the sheets and the stacks of the sheets. While graphite has a single plane of atoms per sheet, in $MoS_2$ each sheet consists a plane of Mo atoms sandwiched between two planes of S atoms. The single-sheet structure can feature trigonal prismatic coordination around Mo, which is the semiconducting 1H ("hexagonal") structure, or octahedral bonding, which is the metallic 1T ("trigonal") phase. 1H and the ideal 1T each have 3 atoms ($MoS_2$) per unit cell (Figure 1); however, theoretical calculations with density-functional theory (DFT)[19] and X-ray diffraction (XRD) experiments [20] have shown that in fact 1T is unstable with respect to distortions. The most commonly reported structure is a $\sqrt{3}\times\sqrt{3}$ distortion that triples the unit cell, known as the 1T' phase [21].

Each single-sheet structure can be stacked into a crystal where the next sheet is exactly above the preceding one (AA stacking), producing the 1H, 1T, and 1T' polytypes. The naturally occurring polytypes in molybdenite, however, are only based on the 1H sheet, and show more complicated stacking [9]. The more common is the 2H structure with 2 sheets per cell in AB stacking, where an S atom in one layer is above an Mo atom in the layer below [22,23]. The less common 3R ("rhombohedral") structure has 3 sheets per cell with ABC stacking [22,24]. These structures are shown in Figure 1. Mo-S bond lengths are around 2.4 Å for all polytypes [25]. DFT calculations show only very small energy differences between the 2H and 3R phases [25]. This insensitivity to stacking, and the low surface energy for 2H-$MoS_2$ of 47 mJ/m$^2$ = 0.025 eV/cell [26], are indicative of low barriers to sliding and hence low friction. By contrast, 1T and 1T' are calculated to be higher in energy by 0.8 eV[25] and 0.5 eV per $MoS_2$ unit, respectively [19].

Lattice parameters and space groups for the polytypes are summarized in Table 1. Nonetheless, deviations from the pristine bulk crystal are common. Calculations have indicated that S vacancies are the most favorable defects in this material [27]. Various kinds of $MoS_2$ nanostructures can be produced, including nanoclusters [28], nanometer-size sheets [29], nanotubes and even inorganic fullerenes [30]. The nanoparticles often show relatively poor crystallinity in XRD [31]. Amorphous or nanocrystalline $MoS_2$ has also been reported [32].

While the 2H and 3R polytypes of $MoS_2$ are found in abundance in nature, their bulk single crystals can also be synthesized in the laboratory using the method of chemical vapor transport (CVT), whereby a two-zone quartz furnace is utilized for a duration of multiple days to continuously vaporize parent elemental solids (in this case, Mo and S) in the presence of a catalytic transport agent (mostly halogens), which eventually results in the synthesis of bulk $MoS_2$ single crystals [33]. Selective synthesis of the 2H and 3R polytypes in CVT can be achieved by the proper choice of transport agent ($I_2$ for 2H, $Cl_2$ for 3R) [33]. Despite the fact that the thermodynamically-favored polytypes of $MoS_2$ (2H and 3R) have been known for a long time, the metastable 1T polytype was only discovered in the early 1990s, through hydration and subsequent oxidation of alkali-intercalated $MoS_2$ compounds (most commonly $KMoS_2$) [21].

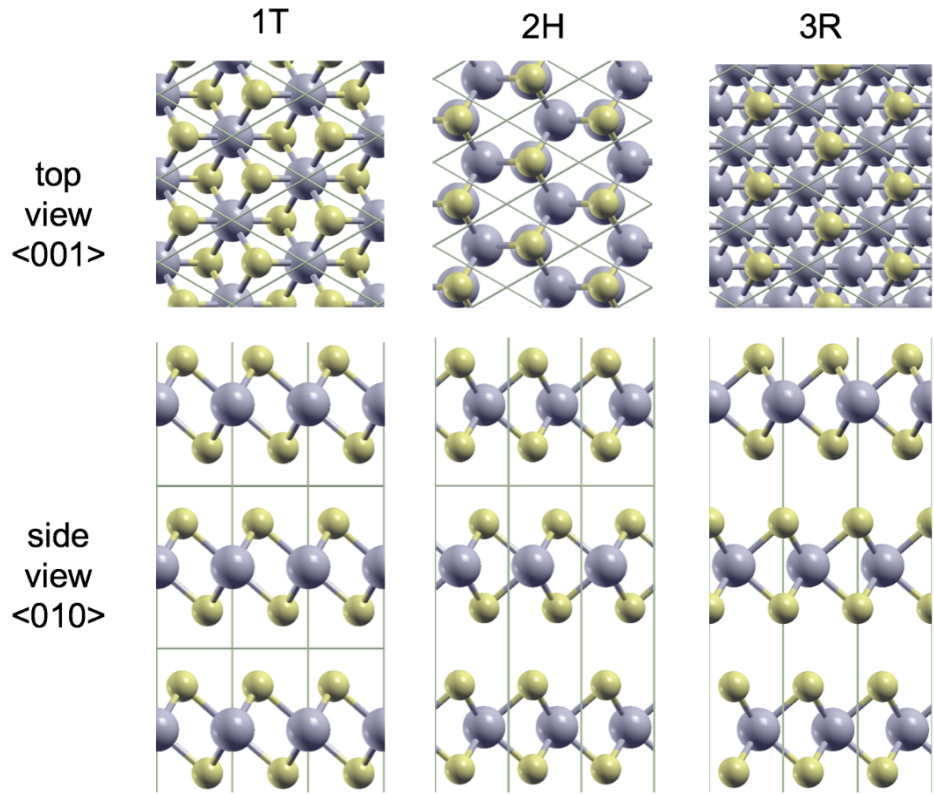

**Figure 1.** Structures of the polytypes of $MoS_2$, constructed from the experimental literature, with the conventional unit cell marked for each. For simplicity, the ideal 1T rather than the distorted 1T' is shown. Mo is grey and S is yellow.

**Table 1.** The structures of the three polytypes of $MoS_2$. In each case, $a = b$, and the unit cell angles are $\alpha = 90º$, $\beta = 90º$, $\gamma = 120º$. Lattice parameters are reported from XRD studies. The 3R structure also has a rhombohedral primitive cell with 3 atoms.

| Polytype | Space group | Point group | Atoms per conv. cell | Stacking | XRD lattice parameters | Properties |
|---|---|---|---|---|---|---|
| 1T' | $P\bar{3}m1$ | $D_{3d}$ | 9 | AAAAAA | $a$ = 5.60 Å, $c$ = 5.99 Å [Ref. [21]] | metallic, metastable |
| 2H | $P6_3/mmc$ | $D_{6h}$ | 6 | ABABAB | $a$ = 3.16 Å, $c$ = 12.29 Å [Ref. [34]] | semiconducting, naturally occurring |
| 3R | $R3m$ | $C_{3v}$ | 9 | ABCABC | $a$ = 3.17 Å, $c$ = 18.38 Å [Ref. [34]] | semiconducting, naturally occurring |

Research on 2D materials has accelerated rapidly in the last decade and a half, thanks to the advent of graphene and the discovery of its exceptional physical properties [35]. Since $MoS_2$ is one of the layered TMDs, its synthesis/production in 2D form (consisting of single or only a few layers) has also been a subject of intense research. While mechanical exfoliation (i.e. cleaving) through the now-famous adhesive tape method [36] or chemical exfoliation via the use of appropriate solvents [37] and electrochemically-induced bubbling [38] regularly yield 2D $MoS_2$ flakes of exceptional quality, their lateral size is only limited to a few tens of micrometers and as such, require the establishment of alternative methods for the synthesis of large-scale 2D $MoS_2$ films for technological applications. The most promising method toward that direction is chemical vapor deposition (CVD), which entails the use of precursor gases flowing over a substrate (e.g. $SiO_2$ [39-41], sapphire [42] or Au [43]) at elevated temperatures in a furnace, which eventually results in the synthesis of 2D MoS$_2$ sheets. With proper choice of precursor gases and deposition parameters, the method now allows the wafer-scale growth and subsequent transfer of high-quality monolayer MoS$_2$ films for use in various promising applications [44]. Another route toward obtaining 2D films of MoS$_2$ involves thermal decomposition (i.e. *thermolysis*) of ammonium thiomolybdate ((NH$_4$)$_2$MoS$_4$), which has been demonstrated at the wafer-scale on insulating substrates, with potential use in device applications [45-47].

While 2D MoS$_2$ films are of interest from a fundamental point of view (e.g. as potential solid lubricants for nano and micro-scale mechanical systems to be employed in space), virtually all current applications where MoS$_2$ is utilized as a solid lubricant involve films / coatings that are significantly thicker (from a few tenths of μm up to a few μm). Traditional methods of applying MoS$_2$ as a solid lubricant onto solid surfaces include approaches such as burnishing and spray bonding; on the other hand, the great majority of MoS$_2$ films employed in modern systems are coated onto target surfaces using physical vapor deposition (PVD) and in particular, *sputtering*. Sputtering involves the bombardment of a MoS$_2$ target by a noble gas plasma in a vacuum chamber, followed by the ejection and eventual deposition of MoS$_2$ particles onto the target substrate, creating a conformal coating. The main advantages associated with sputtered MoS$_2$ films when compared with those obtained by other methods are improved adhesion to the substrate, a higher density and significantly higher purity (owing to the deposition process taking place under vacuum), all of which are factors contributing to improved tribological properties including low coefficient of friction and enhanced wear resistance [8]. Sputtered MoS$_2$ films, which may, in the as-sputtered state, consist of mainly columnar (with MoS$_2$ crystallites oriented perpendicular to the substrate) or basal (with MoS$_2$ crystallites oriented parallel to the substrate) morphologies [48], are utilized both for fundamental studies [49] as well as critical applications, e.g. as solid lubricants on bearings employed in spacecraft [7,50]. A particular advantage of the sputtering method is that other materials can be co-sputtered with MoS$_2$ relatively easily in order to achieve doped coatings with improved properties [15], as discussed further in Section 7. The method of sputtering MoS$_2$ onto the target surface (e.g. RF (radio frequency) based, DC (direct current) based or their magnetron-assisted versions) can also have important implications for the structure and tribological performance of the resulting films [15]. Finally, it should be mentioned that alternative methods such as pulsed laser deposition [51] have also been utilized for MoS$_2$ coatings, although not as widely as sputtering.

## 3. Tribological Applications of MoS$_2$

Generally, solid lubricants are used when liquid lubricants do not meet the advanced requirements of a given application. For example, oils or greases cannot be used in many applications because of issues with application, sealing problems, weight, or environmental conditions [52]. Solid lubricants also reduce weight, simplify lubrication, and in some cases, are less expensive than oil and grease lubrication systems [52]. Table 2 summarizes the conditions and corresponding applications where solid lubricants can be used to overcome limitations associated with standard greases and oils.

**Table 2.** Examples of conditions or environments where liquid lubricants are either undesirable or ineffective and solid lubricants can be used, and representative applications where those conditions exist; adapted from Ref. [52].

| | Space Mechanisms | Medical or Dental Equipment | Nuclear Reactors | Food Processing Equipment | Hard Disks, Microscopes, Cameras | Semiconductor Manufacturing | Furnaces / Metalworking Equipment | Refrigeration / Liquid Ni Pumps | Bridge, Plant or Building Supports |
|---|---|---|---|---|---|---|---|---|---|
| High Temperature | x | | | | | | x | | |
| Cryogenic Temperature | x | | | | | | | x | |
| Radiation | x | | x | | | | | | |
| Corrosive Gas | x | | | | | x | | | |
| High Pressure/Load | x | | | | | | | x | x |
| Product Contamination Unacceptable | | x | | x | x | x | | | |
| Service Difficult or Impossible | x | | x | | | | | | |
| Weight Limited Applications | x | | | | | | | | |

In high or ultrahigh vacuum conditions, liquid lubricants can evaporate such that they are unable to perform their lubricating function and can potentially contaminate the device. Similarly, liquid lubricants can decompose or oxidize at high temperatures. At the other end of the spectrum, at cryogenic temperatures, liquid lubricants can solidify or become too viscous to flow effectively. Under radiation or corrosive gas environments, liquid lubricants can decompose. High pressure applications can exceed the load carrying capacity of liquids. In cases where contamination is a major issue, liquid lubricants can be problematic because they tend to pick up dust and other contaminants from the environment. In applications where weight is of primary importance, liquid lubricants and the associated components can be too heavy. Lastly, if service is difficult or impossible, it may not be viable to use a liquid lubricant that degrades over time, particularly if the application experiences long storage or idle periods. Solid lubricants overcome many of these issues.

Notable in Table 2 is one application where almost all the conditions for poor liquid lubrication exist, i.e. in space mechanisms. Moving systems in space include vehicles, satellites, telescopes, antennas and rovers. Such systems must operate for long periods of time with little or no service under a wide range of extreme environmental conditions. Since most space applications operate in vacuum environments, the typical solid lubricant of choice is $MoS_2$ [7,53]. Unlike graphite, which requires sufficient vapor pressure of water to provide lubrication, $MoS_2$ performs best in vacuum environments [54]. Representative examples of components in space applications that already rely on $MoS_2$ lubricants are ball bearings, pointing mechanisms, slip rings, gears and release mechanisms [15,55]. It has indeed been suggested that most if not all the American satellites and spacecraft contain $MoS_2$ for some application [13].

The continuing interest in the further development of solid lubricants (including $MoS_2$) for space applications stems from the fact that planned space missions increasingly involve a wide range of operating conditions (in particular, in terms of temperature) and increasing durations during which the mechanisms that enable relative motion will need to operate in a robust and reliable fashion without the means of regular maintenance. There is perhaps no better example than the catastrophic high-gain antenna deployment failure associated with the Galileo spacecraft [56], caused by the failure of the MoS$_2$-based solid lubricant during operation in space that was successfully tested before deployment on earth under ambient conditions. As such, basic and applied research is being conducted around the world to understand and improve the properties of solid lubricants in general and MoS$_2$-based lubricants in particular, for meeting the requirements of demanding space missions. Illustrative examples of such ambitious missions include NASA's planned lander mission to Europa for astrobiology research as well as ESA's BepiColombo mission to Mercury, where the spacecraft will be exposed to temperatures on the order of 250 °C [7].

Based on the fact that MoS$_2$ is the most extensively used solid lubricant in space mechanisms, a particularly timely example will be highlighted here: the James Webb Space Telescope (JWST), which is planned for launch in 2021 and is the successor to the Hubble Space Telescope. MoS$_2$ has been determined as the solid lubricant of choice for sensitive mechanisms employed in a number of precision instruments on JWST, including the Near Infrared Spectrograph (NIRSpec) and the Mid Infrared Instrument (MIRI) [50,57]. Critical to the choice of MoS$_2$ as the preferred solid lubricant for these high-precision instruments was the fact that it preserves its lubricative properties down to cryogenic temperatures (which include 30 K, the temperature around which JWST will be operating)[58] and the high uniformity with which it can be deposited on precision bearings employed in the instruments at sub-micron thicknesses, ensuring stable operation. In order to minimize the chances of a lubricant-related catastrophic failure after deployment (like the Galileo), a significant amount of work has been performed to characterize the tribological performance of bearings under various operating conditions on earth and it has been determined that a run-in procedure performed under vacuum is necessary to ensure low friction torques for eventual use in space [59].

The impressive tribological properties of MoS$_2$ under vacuum and in a wide range of temperatures make it a natural choice as a solid lubricant for applications in space; however, the fact that its lubricity rapidly deteriorates under humid conditions represents a significant barrier to its extensive use in demanding terrestrial applications. On the other hand, recent developments in coating technology, in particular the use of closed field unbalanced magnetron sputter ion plating (CFUBMSIP) employed to co-sputter MoS$_2$ with Ti, has resulted in lubricant coatings with notable improvement in mechanical properties and resistance to humidity [60-62]. These improvements resulted in the successful use of MoS$_2$-based coatings in dry machining operations such as cutting and forming, as well as certain machine components that do not operate under vacuum conditions. Despite these developments, it is projected that the main use of MoS$_2$ as solid lubricant will be in aerospace/space applications for the foreseeable future, with ongoing fundamental and applied research directed at adjusting the properties of MoS$_2$ coatings toward long-term missions and a wider range of operating conditions.

Before concluding this section, it needs to be mentioned that MoS$_2$ has numerous other applications besides lubrication that are not covered in this review. One of the most important is heterogeneous catalysis, in which active sites are edges, defects, or near "promoter" dopants (Ni, Co). MoS$_2$ can catalyze reactions including hydrodesulfurization, hydrogen evolution reaction, oxygen reduction reaction and photo-catalytic water splitting [63]. The ability of MoS$_2$ to absorb Li between its layers (intercalation) has been utilized for batteries [63]. Electronic and optical applications have become a particular focus after the advent of few-layer MoS$_2$, which have been reviewed recently [18,64]. Some unique features are strong excitonic effects, a direct-to-indirect band-gap transition with number of layers, tunability by strain, and ease of creating heterojunctions by stacking. The presence of multiple inequivalent low-energy valleys in the band structure might be used in "valleytronics," and the strong spin-orbit coupling suggests spintronic applications. These properties have been explored for use in transistors, sensors, photovoltaics, light-emitting diodes, and lasers [18,64].

**4. Mechanisms of Low Friction and Wear**

While MoS$_2$ has been known to be a good solid lubricant for many years [12], the mechanisms through which this performance is achieved have been a subject of ongoing research. This section will provide an overview of experimental and computational efforts aimed at understanding the physical reasons behind the extraordinary lubricity of this lamellar material. Moreover, key results of atomic force microscopy (AFM) based friction studies on 2D $MoS_2$ samples (which only consist of a single layer or a few layers stacked on top of each other) will be reviewed.

The most widely-referenced works in the literature aimed at elucidating the lubricative properties of $MoS_2$ are the detailed transmission electron microscopy (TEM) studies by Martin *et al.* that were performed in the early 1990s [49,65]. While techniques such as XRD have been employed before to understand some aspects of the friction mechanisms related to $MoS_2$ coatings [66], the TEM images provided convincing visual evidence for the processes by which $MoS_2$ was thought to exhibit low friction and wear [49,65]. In particular, by performing pin-on-flat tribometer experiments under ultrahigh vacuum conditions on sputter-deposited, stoichiometric, polycrystalline $MoS_2$ films followed by TEM imaging of the wear debris as well as the film itself, the following key processes were determined to take place during relative sliding motion between components lubricated by $MoS_2$:

(1) The establishment of a *transfer film* on the counter-surface as it slides against the $MoS_2$-coated component.

(2) The shear-induced orientation of the basal planes of $MoS_2$ (in the original coating, the transfer film and eventually in third bodies / wear particles) in the sliding direction.

A key feature that leads to ultra-low friction characteristics (mostly referred to as *superlubricity* [67], with friction coefficients on the order of 0.001) of $MoS_2$ coatings has been determined, via a careful study of the Moiré patterns that emerge during TEM imaging, as the fact that there is significant variability in the rotational registry of individual $MoS_2$ crystallites around the surface-normal direction. This leads to a rotational structural mismatch between basal plane pairs that come into contact with each other at the interfaces that form between the original coating, the transfer film and the third bodies, leading to a systematic cancellation of atomic-scale lateral forces. This results in the establishment of an ultra-low friction regime, in agreement with the theory of *structural superlubricity* that was spearheaded by Hirano [68] and Sokoloff [69], and is still being investigated in many different experimental systems [67].

Even though the pioneering TEM works described above pinpoint structural superlubricity between basal planes of $MoS_2$ as the main reason for its ultra-low friction characteristics, a particular limitation involves the inability to directly observe, in an *in situ* fashion, the sliding process between the layers and to directly measure the shear forces that take place during sliding. These shortcomings were partially addressed in a careful experimental study performed by Oviedo *et al.*, where two custom cross-sectional TEM setups were used to (i) visualize *in situ* with high-resolution the sliding between individual basal planes of $MoS_2$ and (ii) quantify the shear strength of the interface between the layers [70]. In particular, the upper bound for the shear strength (by focusing on interlayer sliding between $MoS_2$ layers in commensurate registry) was determined to be 24.8 ±0.5 MPa. In another important work, utilizing a mechanical force sensor in the form of a silicon nanowire inside a scanning electron microscope (SEM), an individual, single-layer flake of $MoS_2$ was controllably detached from and slid on its bulk parent $MoS_2$ substrate in structurally incommensurate registry [71]. In what constituted the first direct measurement of friction forces between incommensurate basal planes of $MoS_2$, the authors obtained an ultra-low friction coefficient on the order of 0.0001 (well within the superlubric regime [72]), supporting the original proposition by Martin *et al.* that easy-shear sliding between incommensurate basal planes of $MoS_2$ are responsible for its ultra-low friction. In yet another rather challenging experimental study, AFM tip apexes were intentionally wrapped with sheets of 2D materials (including $MoS_2$) and slid on a bulk graphite substrate, with experimentally recorded friction coefficients smaller than 0.002 [73].

Due to experimental difficulties associated with the measurement of friction forces between individual $MoS_2$ planes in an isolated fashion and the requirement of highly customized setups that are not readily accessible, computational efforts play an important role in the elucidation of low friction mechanisms of $MoS_2$. Importantly, simulations of sliding between individual $MoS_2$ sheets based on molecular dynamics (MD) have uncovered (i) a 100-fold decrease in friction force when the sheets are rotated with respect to each other to transition from a commensurate to an incommensurate state, (ii) that the highest energy barriers to sliding are located on top of S atoms and (iii) that Coulombic repulsion between the layers plays an important role in the easy shear characteristics, supplementing the reduction in energy barriers to sliding that is induced by structural mismatch [74,75]. A subsequent study based on DFT focused on the load-dependence of friction between individual $MoS_2$ planes by simulating potential energy landscapes, and determined that inter-layer friction increases with increasing load based on electrostatic effects [76]. Another study of inter-layer motion using both DFT and MD showed that the energy barriers to sliding in $MoS_2$ are larger than those of other 2D materials (graphene and boron-nitride) due to the corrugated nature of the $MoS_2$ surface, which increases both the dispersion binding and the energy barriers, and also because the sulfur anions are more polarizable than the first-row atoms in graphene and boron-nitride [77].

The main characteristics that make $MoS_2$ a low-friction material (in particular, the ability of its basal planes to orient in the direction of sliding) are also mainly responsible for its wear-resistant properties. TEM-based studies performed on $MoS_2$ films deposited by CVD have established that mechanical properties including strength, elastic modulus and strain to fracture improve during the initial stages of sliding, accompanied by a marked decrease in wear rate [78]. The physical mechanism behind the *in situ* enhancement of mechanical properties and thus, wear resistance, is thought to involve an interplay of film densification (induced by individual crystallites pushing into the pores in the as-synthesized films under contact stresses) and the ability of the $MoS_2$ crystallites to easily re-orient themselves in the direction of sliding to accommodate shear strain and delay the onset of fracture that would eventually manifest as wear [78,79]. The wear resistivity of pure $MoS_2$ films can be further augmented by the use of dopants [80], as described in Section 7.

The invention of the AFM [81] and the realization that it can be used to study frictional properties on the nanometer scale [82], opened up a new era in tribology research where friction could, for the first time, be studied at the "single asperity" level, with the potential to provide fundamental physical information about this ubiquitous phenomenon. Moreover, the realization that graphene can be readily obtained from its lamellar bulk form by simple mechanical exfoliation [83] and the booming interest in related 2D materials that followed, resulted in a number of AFM-based studies regarding the frictional properties of such materials. The motivation to study the frictional properties of 2D materials is mainly based on the desire to establish robust and effective lubrication schemes for nano- and micro-scale mechanical systems, where increasing surface-to-volume ratios amplify surface effects (e.g. surface tension, stiction) that preclude the use of liquid lubricants and lubrication via thick layers of lamellar solids is generally impractical because of geometric constraints associated with the size of the components.

At this point, it should be emphasized that AFM-based friction studies on 2D materials (including $MoS_2$) probe mechanisms and properties that are fundamentally different from those that are largely responsible for the lubricity of such materials in their bulk form. In particular, while the excellent frictional properties of $MoS_2$ have been mainly attributed to easy shear between its basal planes and related processes as discussed above, no interlayer shear takes place during AFM experiments, where the measured friction forces occur at the interface between the AFM tip and the top surface of the $MoS_2$ sample. Despite this fact, it is perhaps surprising that AFM-based studies of friction on such materials still reveal miniscule friction forces and drastically decreased friction when compared with typical substrates (e.g. $SiO_2$) on which they are deposited. As no interlayer sliding takes place during experiments that feature a few-layers of material (and interlayer sliding is, by definition, not possible for single-layer samples), the significant lubricative properties of 2D materials in AFM experiments are attributed to other physical factors [84], including (i) their atomic-scale smoothness, (ii) mechanical strength and (iii) chemical inertness.

Despite the fact that the great majority of AFM-based nanotribological studies on 2D materials have been performed on graphene, single- or few-layer $MoS_2$ have also been the target of some recent experimental work. Although AFM-based friction measurements on single-layer $MoS_2$ have been performed as early as 1996 [85], the first systematic, comparative report of AFM-based friction measurements on 2D materials (which also includes results obtained on mechanically exfoliated flakes of single- and few-layers of $MoS_2$) appeared in 2010 [86]. Specifically, it was observed that the magnitude of friction measured on $MoS_2$ (deposited on $SiO_2$ substrates) decreased monotonically with increasing number of layers (until a total of ~5 layers was reached). The authors attributed the consistent behavior of decreasing friction with increasing number of layers to a mechanical process called *puckering*, whereby the AFM tip apex that probes the 2D samples creates locally deformed areas (*puckers*) around itself, which increases friction due to an increase in contact area. As the number of layers increases, the bending stiffness of the 2D material in the vertical direction also increases, leading to a decrease in puckering and consequently, in friction. While factors such as electron-phonon coupling [87] as well as surface roughness [88] may also play a role in the layer-dependent friction of 2D materials, the puckering phenomenon is generally accepted to be the dominant mechanism.

Recent work performed on polycrystalline, single- to few-layer $MoS_2$ samples grown by CVD featuring grains up to a few hundreds of nanometers in size revealed a strikingly different layer-dependence of friction than single crystal samples obtained by mechanical exfoliation [89]. In particular, under ambient conditions, the layer dependence was found to be of an "oscillatory" nature, whereby samples with odd numbers of layers exhibited higher friction relative to those samples with even numbers of layers. This rather interesting result was attributed to a mechanism involving the enhanced adsorption of charged species at samples with odd numbers of layers due to the existence of permanent dipoles. The adsorption of such species change the interactions of the sample surface with the probing tip apex, ultimately leading to enhanced friction when compared with samples featuring even numbers of layers [89].

In AFM experiments, it is widely observed that the physical properties of the tip apex have a profound effect on the recorded data. This observation is also true for AFM-based investigation of friction mechanisms on $MoS_2$, as it has been recently shown that layer-dependent friction on 2D $MoS_2$ becomes strongly non-monotonic (first decreasing when switching from one to two layers, and then progressively *increasing* with increasing number of layers) when probed with blunt, pre-worn tips [90]. These results point to the conclusion that the physical properties of individual sliding components, in addition to those of the solid lubricant, need to be carefully evaluated in the design of small-scale mechanical systems to achieve adequate lubricative performance.

The *anisotropy* of friction on $MoS_2$ is another topic that has been investigated via AFM on 2D samples. Specifically, Cao *et al.* determined, by means of consecutive friction measurements performed on $MoS_2$ flakes gradually rotated with respect to the scanning direction during AFM experiments, that the friction forces on $MoS_2$ exhibit anisotropic behavior with a periodicity of $180°$ [91]. This observation is surprising at first glance, as a periodicity of $60°$ would have been expected based on crystal symmetry. Inspired by previous anisotropy work on graphene [92], the authors attributed their observations to the existence of parallel *ripples* in the $MoS_2$ samples induced by in-plane strain that occurs during the deposition of the samples on the $SiO_2$ substrate in conjunction with the puckering effect, which leads to the emergence of high- and a low-friction sliding directions (separated by $90°$) that correspond to the tip moving across or along the ripples, respectively. While the work by Cao *et al.* also discusses the effect of sample thickness and load on friction anisotropy in a convincing fashion, the specific role played by the AFM tip apex (for instance, by its size and stiffness) in friction anisotropy remains to be studied in future work.

The frictional properties of 2D $MoS_2$ have been recently compared with those of graphene [93], employing samples grown by CVD, which provides films of lateral sizes that are much more suitable for practical purposes than those obtained by mechanical exfoliation. By combining AFM-based imaging and micro-tribometer tests, it was determined that the $MoS_2$ film exhibited superior friction and wear characteristics when compared with graphene. The authors attributed this observation to the fact that $MoS_2$ is directly grown on the $SiO_2$ substrate employed in the measurements, whereas graphene needs to be grown on copper foils first and then transferred onto $SiO_2$ through chemical means. This process inevitably introduces contaminants, which prevent intimate contact and adhesion between graphene and the underlying substrate, ultimately leading to inferior friction and wear performance [93], This result is in contrast to another recent direct comparison between MoS$_2$ and graphene conducted via AFM measurements of both materials in the same scan line that showed graphene consistently exhibited lower friction. The experimental results were analyzed using DFT and MD simulations which revealed the mechanism underlying the friction contrast to be a higher energy barrier to sliding on MoS$_2$ [94]. These studies will almost certainly encourage future comparisons between 2D materials which will highlight their strengths and weaknesses and inform their use in specific applications.

## 5. Environmental dependence

The tribological behavior of MoS$_2$ is extremely sensitive to environmental conditions. As stated before, one of the main prerequisites to achieve ultralow friction with MoS$_2$ lubricants is the absence of contaminants such as oxygen, water and hydrocarbons [49]. Among the first observations of the environmental dependence of the lubricative properties of MoS$_2$, Peterson and Johnson[95] showed in 1953 that the friction between metal surfaces lubricated by MoS$_2$ increased with relative humidity (R.H.) up to 65% and then decreased. After this work, the presence of water and various forms of oxygen has been shown to increase friction and wear in several studies [96-100]. In the following, we summarize the effect of oxygen and water on the friction and wear behavior of MoS$_2$ as a solid lubricant.

The presence of molecular and atomic oxygen can lead to surface-limited oxidation (or near-surface-limited, due to the subsequent protection that occurs via the oxidation layer) of MoS$_2$ films, significantly affecting their tribological performance. The oxidized layers of MoS$_2$ films can cause high friction during the run-in process; then, once the initial oxide has been removed, the friction drops to its low steady-state value as long as the oxide removal rate is higher than the oxidation rate in the system [101,102]. This high friction run-in regime can cause problems specifically in aerospace and satellite devices in which operation is infrequent and oxidation can occur by molecular oxygen during the assembly and testing of a device on earth [103]. Oxidation may also happen in the presence of atomic oxygen, which is extremely reactive and quickly oxidizes the surface of MoS$_2$ [104], particularly during operation in the low earth orbit (LEO) environment (altitudes of 100−1000 km). This pre-operation exposure to the terrestrial environment before space missions enhances the premature failure risk in satellite devices. The failure of the high gain antenna in the Galileo spacecraft is an example caused by this problem, whereby the failure of bearings during deployment was attributed to the prolonged storage and transportation of the components [105].

Molecular oxygen was shown not to affect the tribological properties of MoS$_2$ at room temperature [101,106]; however, it can oxidize the MoS$_2$ edge sites at high temperature, whereby MoO$_3$ (and MoO$_2$), as a product of this oxidation, disrupts the easy shear properties of the MoS$_2$ lubricant film and creates high wear [107]. The oxidation of MoS$_2$ starts with the physisorption of oxygen on the surface of MoS$_2$ and then the chemical formation of Mo oxides. MoS$_2$ oxidation increases with an increase in temperature [108]. The transition temperature at which the oxidation rate increases substantially has been reported to be from 100 °C [109,110] to 400 °C [111,112], which may depend strongly on the preparation method of the MoS$_2$ film and on *film conditioning* induced by deliberate thermal cycling during sliding prior to testing [101]. In general, powdered coatings (i.e. mechanically deposited coatings) exhibit higher transition temperatures compared to sputtered coatings, which might stem (i) from larger grain sizes in the mechanically deposited films which decrease the susceptibility to oxidation and/or (ii) the porous structure of the sputtered films which can speed up the oxidation of MoS$_2$ as a result of tribochemical reactions [101,111,113,114]. The effect of microstructure on tribological performance of MoS$_2$ coatings under ambient environment was further studied by comparing the friction of highly oriented N$_2$-spray-deposited MoS$_2$ films and amorphous films produced via DC magnetron sputtering [115,116]. Highly ordered MoS$_2$ films with surface-parallel basal orientation exhibited higher resistance to oxidation compared to amorphous MoS$_2$ films. Moreover, X-ray photoelectron spectroscopy (XPS), and high-sensitivity low-energy ion scattering (HS-LEIS) analysis revealed that highly oriented MoS$_2$ lamellae restricted the oxidation occurrence to the first top few layers, which led to shorter run-in period compared to amorphous

MoS$_2$ in which the further penetration of oxygen into the surface resulted in a longer run-in period [116]. Therefore, deposition techniques that result in a lower density of edge sites or highly oriented lamellae of MoS$_2$ can reduce the possibility of oxidation, thereby minimizing the degradation of tribological performance under ambient conditions.

Atomic oxygen (AO) is an abundant species (with a high flux density on the order of $10^{13}$-$10^{15}$ atoms·cm$^{-2}$·s$^{-1}$ and kinetic energy ~5 eV) that acts on exposed surfaces of devices and components employed in space applications, and can cause deterioration of material properties [117]. There has been extensive research on the effect of AO on the tribological performance of MoS$_2$ lubricants (see, e.g. [114,117-121]). However, there are some discrepancies in the reported results: for instance, no change in the tribological properties of MoS$_2$ after AO exposure was reported by Ref. [120]; whereas most other studies reported high initial friction due to the limited depth of oxidation, and S loss on the MoS$_2$ coating surface [117-119,121]. Moreover, Wang *et al*. observed an overall higher mean friction for a pure MoS$_2$ sample with AO exposure as well as a direct correlation between the wear-volume/friction and the AO exposure time [114]. The atomic oxygen flux density during sliding has also been shown to inversely affect the life of the coating [119]. The discrepancies in these results might stem from differences in experimental conditions such as atomic oxygen beam energy (thermal to keV), fluence (i.e. intensity, $10^{12}$–$10^{24}$ atoms/cm$^2$), sample preparation and/or tribological test conditions [122].

The effect of a humid environment on the friction and wear properties of MoS$_2$ lubricant coatings has been extensively studied since the mid-20$^{th}$ century [12,95,98,99,101,123]. Generally, the presence of water molecules in the environment (e.g. while storing MoS$_2$ under humid conditions) leads to an increase in sliding friction and wear rate regardless of the MoS$_2$ film deposition technique employed [13,85,97,124,125]. Despite general agreement in the literature about the effect of water on the tribological behavior of MoS$_2$, there are discrepancies regarding the governing mechanism associated with this phenomenon. Common hypotheses for the increase in friction in humid environments include:

- Water-driven oxidation of MoS$_2$ edge sites [96-99,124],

- Physisorption of water molecules at the surface which deteriorates the tribological behavior of MoS$_2$ via disruption of the easy shear of lamellae [123,125-129], adhesion enhancement [130,131], hydrogen bonding between basal planes of MoS$_2$ [132,133] and restriction of the growth and reorientation of the tribofilm [115].

For many years, oxidation of MoS$_2$ at edge sites was thought to be the mechanism responsible for increasing friction and wear rate of MoS$_2$ under humid environments [96-99,124]. This hypothesis was mainly tested in the presence of both water and oxygen species and an enhancement in oxide formation (and subsequent sulfur loss at the surface) was observed as the relative humidity (i.e. partial pressure of water) increased. As it was mentioned before, oxygen alone can oxidize the MoS$_2$ film, so the important question about the effective and distinctive role of water in oxide formation remained unanswered in these studies. Recent experiments [101,109,123,125], and simulations [128] decoupled the role of water and oxygen in the tribological behavior of MoS$_2$ in humid environments. It was found that water did not promote MoS$_2$ oxidation at room temperature. This statement was supported by energy-dispersive X-ray spectroscopy (EDS) analysis in which no oxygen K$α$ was detected on the worn surface of lubricant films in humid nitrogen environment; i.e. water did not cause the degree of oxidation necessary for detection in this test [123]. This finding is in agreement with Raman spectroscopy results of Windom *et al*. in which a humid environment exhibited little effect on the oxidation compared to dry air or O$_2$ environments [109]. Moreover, increasing the temperature to values between room temperature and the transition temperature for oxidation greatly improved the tribological properties of the coating [123,125]. This evidence further highlights the fact that the adsorption/desorption of water is a reversible process, i.e. physisorption rather than chemisorption. Furthermore, *ab initio* MD simulations modelling the interaction of water with MoS$_2$ bilayers showed that intercalated water molecules deteriorate the sliding motion of both regular and defective layers considerably and that the sliding distance and velocity after a given shear stress are reduced by increasing amounts of water molecules at the interface, consistent with viscous friction behavior [128]. In a recent study, Lee *et al.*[129] attributed the friction enhancement in the presence of intercalated water to the enhancement of phononic energy dissipation at the $MoS_2$-water interface. It has also been suggested that liquid water could be formed by capillary condensation of vapor in the defects of the $MoS_2$ crystal structure and that water could then deteriorate the easy shear between basal planes [130]; however, no direct observation of adhesion enhancement has yet been reported.

A new interpretation on the role of water in the tribological behavior of $MoS_2$ was recently proposed by Curry *et al.* [115]. It was observed that the run-in coefficient of friction for highly-oriented $MoS_2$ remains the same under dry and humid environments, whereas the run-in behavior is highly environment-dependent in the case of amorphous films. Consequently, it was suggested that water restricted the formation and growth of the shear-induced, highly ordered tribofilms instead of deteriorating the shear strength of existing highly ordered tribofilms. This hypothesis needs further investigation for confirmation.

Recent computational studies [128,134-136] on the energetics and mechanism of water adsorption and dissociation on $MoS_2$ films showed that oxidation of $MoS_2$ layers by water is much less likely than the simple adsorption of water molecules at edge sites. Additionally, the most favorable locations for the adsorption of water molecules were determined to be Mo edge sites. However, at room temperature, water molecules could be dissociated to O and OH that bind to all available edge sites on $MoS_2$ (i.e. not just Mo) [135]. The higher reactivity of the edges in $MoS_2$ lubricant films emphasizes the potential role of the microstructure in the tribological response of $MoS_2$ under ambient conditions; the presence of highly ordered structures can significantly improve the tribological behavior of $MoS_2$ coatings in humid environments [48,115,137-139]. To tailor the microstructure for improving the lubricative properties of $MoS_2$ under ambient conditions, Chhowalla *et al.* [140] followed a different approach and generated a thin film of lubricant made of hollow fullerene-like $MoS_2$ clusters ("onions") using the localized arc discharge method. Ultra-low friction and wear were observed in a humid environment and attributed to the presence of curved S–Mo–S planes that prevented oxidation and preserved the layered structure [140].

Overall, the presence of any environmental contaminant degrades the excellent tribological properties of $MoS_2$ observed in vacuum. Below the transition temperature for oxidation, water adsorption disrupts the easy-shear properties of $MoS_2$, whereby oxygen plays only a marginal role in degrading its tribological properties. The adsorption of water is thermodynamically favored, and water diffusion in the bulk has a direct correlation with R.H. and exposure time. As the temperature increases, water species desorb and an improvement in the tribological properties of $MoS_2$ can be observed. As the temperature approaches the transition temperature, oxygen-driven, surface-limited oxidation becomes significant and thus, oxidation deteriorates the excellent lubricity of $MoS_2$ at high temperatures in ambient conditions [123].

Research efforts aimed at achieving environmentally-robust $MoS_2$ solid lubricant coatings mainly employed two approaches so far: changing the method and parameters of deposition and adding dopants. First, microstructural properties can be improved by varying the deposition method (such as by using $N_2$-spray-deposition [115], unbalanced magnetron sputtering [137] or ion-beam-assisted deposition [141]) or parameters (such as lowering sputter gas pressure [142]) during film preparation to obtain dense films with highly oriented basal planes parallel to the substrate and with the lowest possible density of defects and edge sites. As briefly mentioned above, this approach can improve the environmental resistance of $MoS_2$ lubricants [115,137-139]. Second, by addition of a small amount of other materials such as metals and inorganic sulfides/oxides, the coating density (and consequently, internal stress), hardness and oxidation resistance can be improved. The effect of dopants on the tribological properties of $MoS_2$ lubricants is extensively discussed in Section 7.

## 6. Temperature dependence

The dependence of the tribological properties of $MoS_2$ on temperature has been the subject of several studies [101,110,143-150], where it has been generally shown that friction and wear of $MoS_2$ vary with temperature. However, there is notable disagreement in these reports for the low temperature tribological behavior of $MoS_2$ (from the cryogenic regime to around -150 °C). Some studies [145,146] reported monotonic increases in friction coefficient with decreasing temperature, while other studies [143,144,147,148] suggested that decreasing the temperature below a certain level does not lead to increasing friction coefficients. At moderate temperatures (up to around 200 °C), the friction, in a thermally activated process, monotonically decreases as temperature increases. Wear rate in this temperature range has been reported to increase as temperature increases, such that the wear rate of an $MoS_2$ coating doubled as temperature increased from –100 °C to 100 °C [143]. Curry *et al.* [149] recently developed a predictive model to correlate the macroscale interfacial shear strength of $MoS_2$ to temperature, based on the temperature-dependent probabilities for commensurate or incommensurate sliding to occur, taking into account the corresponding energy barriers to sliding. One hypothesis that has emerged from these studies is that the thermally activated friction behavior is limited to situations where negligible wear takes place, and that the transition from thermally activated friction to athermal friction is a direct result of wear.

At high service temperatures (i.e. above the transition temperature defined in Section 5), friction increases with increasing temperature; under ambient conditions this is due to increasing oxidation rates, whereas the thermal dissociation rates of $MoS_2$ rapidly increase in inert gas or vacuum environments at temperatures above 500 °C [101,110,150,151]. Within this context, the maximum temperature at which $MoS_2$ provides effective lubrication is highly dependent on (i) the operating conditions, e.g. the presence or absence of vacuum and the humidity level, and (ii) the microstructural properties of the $MoS_2$ coating, including its crystal structure, density, and surface roughness. For instance, in vacuum, the maximum operating temperature of burnished $MoS_2$ is in the range of 600 °C – 700 °C because thermal dissociation of $MoS_2$ around these temperatures deteriorates its lubricative properties [150]. However, mechanically deposited $MoS_2$ coatings under ambient conditions could have a maximum operation temperature of 400 °C due to the rapid oxidation of $MoS_2$ [107,151]. The main challenge therefore is to develop $MoS_2$ based coatings which can be used over a wide range of operating temperatures. In order to extend the operating temperature range of $MoS_2$ coatings, adaptive composite coatings have been widely studied. The composite coating improves lubrication not only by the presence of the coating constituents but also by the lubricious products of chemical or physical reactions that happen during sliding. While a comprehensive coverage of the utilization of $MoS_2$ in composite coatings is not within the scope of this review article, several notable examples will be mentioned here due to their relevance for temperature dependence work. Examples of such adaptive systems include $PbO/MoS_2$ [51,152], $YSZ–Ag–Mo–MoS_2$ [153], $Mo_2N/MoS_2/Ag$ [154] and $Ag/MoS_2/graphite$ [155] which extend effective lubrication capabilities over a wide temperature range by combining the lubricity of $MoS_2$ at lower temperatures and the creation of oxides with easy shear properties at high temperatures [156-158]. However, the chemical reactions are not reversible and the oxides do not provide lubrication at low temperatures, so ability to withstand thermal cycling is a challenge [159]. One approach is to establish barrier layers to protect some $MoS_2$ from oxidation [160]. In a different approach to extend the effective lubrication temperature range, Zhang *et al.*[161] achieved a low friction coefficient (~0.03) and low wear rate (~10−13 $mm^3$/N·mm) at 300 °C using carbon nanotube (CNT) based composite coatings. These improvements in tribological properties were attributed to an enhancement of mechanical strength due to the high load-bearing capacity of CNTs. Several studies [162,163] showed that diamond-like carbon (DLC) can also be added to the adaptive coating in order to enhance its service temperature range. The DLC phase in these coatings increased hardness and provided a source of carbon for surface graphitization effective for humid environment lubrication. For a detailed description of adaptive coatings designed to improve the temperature range for effective lubrication by $MoS_2$, interested readers are referred to Refs. [156,164].

## 7. Structure and Tribological Properties of Doped $MoS_2$

### 7.1 Structure of Doped $MoS_2$

Most studies of doped $MoS_2$ have focused on synthetic approaches and effects on macroscopic properties such as electronic structure or catalysis. Here we review the available literature with atomistic characterization of the structures of doped $MoS_2$, with a focus on bulk rather than monolayers. Experimental characterization of a heterogeneous structure (unlike a single crystal) is quite challenging, and therefore DFT calculations have played a key role in elucidating structures.

Dopants in $MoS_2$ can be incorporated in several possible locations. Substitution can occur at either the Mo or S site. Atoms can be added (intercalated) between the $MoS_2$ layers. They can also be adsorbed at a surface, as adatoms on an $MoS_2$ sheet (basal plane) or on an edge, which is particularly likely for few-layer $MoS_2$ or nanosheets. In principle, there could be interstitials within an $MoS_2$ layer, in the sites within an Mo plane, but there is very limited space and such a structure does not seem to be observed. Adatoms can occur in several high-symmetry sites such as atop Mo, atop S, bridging an Mo-S bond, or in the center of a hexagonal hollow [165]. If an atom is intercalated between two layers in the 2H polytype, these become just two high-symmetry positions: tetrahedrally or octahedrally coordinated [166]. These possible dopant locations in bulk 2H $MoS_2$ are depicted in Figure 2.

The favorable dopants and their locations can be understood with several considerations [167]. Since $MoS_2$ formally involves $Mo^{4+}$ and $S^{2-}$ ions, atoms that can adopt a 4+ oxidation state naturally fit the Mo site, while atoms that can adopt a 2- oxidation state fit the S site. Intercalated atoms are expected to be neutral. The atom's ionic radius and its compatibility with the Mo or S radii, whether the atom has similar bond length with Mo or S when compared to those in $MoS_2$, and whether the atom forms a crystal with Mo or S of similar structure to $MoS_2$ are key points. When these are all similar to $MoS_2$, alloying in varying proportions may be possible, as is the case for $Mo_{1-x}W_xS_2$, $MoS_{2(1-x)}Se_{2x}$, and $Mo_{1-x}Nb_xS_2$, which have favorable mixing energies and can show substitutional disorder [166,168]. The chemical potentials of Mo, S, and the dopant under synthesis conditions will control which dopant location is favored: for example, S-rich conditions favor Mo substitution [169]. Doping can perturb the local structure (including symmetry breaking and Jahn-Teller distortions) [170,171], alter lattice parameters, or even favor an overall change in crystal structure to a different polytype: Li intercalation causes a change from 2H to 1T [172], and Nb substitution of Mo favors 3R over 2H [173]. Dopants can affect the growth process [174] and the morphology of the resulting crystals or nanostructures, via kinetic or thermodynamic effects [167]. Dopants can be well dispersed or may tend to cluster [169], and in some cases may segregate into competing phases [174,175].Theoretical studies are complicated by the non-equilibrium conditions in many synthesis approaches, meaning that not only lowest energy structures but also other thermodynamically metastable phases can be formed.

Most doping studies of $MoS_2$ have used transition metals, which are chemically similar to Mo. DFT calculations for monolayer $MoS_2$ found that Fe, Mn, W, Cr, V, and Ti dopants substituting for Mo maintain its six-fold coordination, whereas Zn, Au, Ag, Cu, Pt, Pd, Ni, and Co break symmetry and adopt 4-fold coordination [170]. A DFT study of Mo substitution on the basal plane surface of the bulk found 5-fold coordination for Co, Ni, and Cu, and a 6-fold coordination but distorted geometry for Fe; all had shorter M-S bond lengths. Substitution at the edge rather the basal plane was energetically favored [171]. By contrast to the DFT study, experimental results on Zn doping of $MoS_2$ nanosheets showed little change in Raman or XRD, and were interpreted as Mo substitution without structural distortion [176].

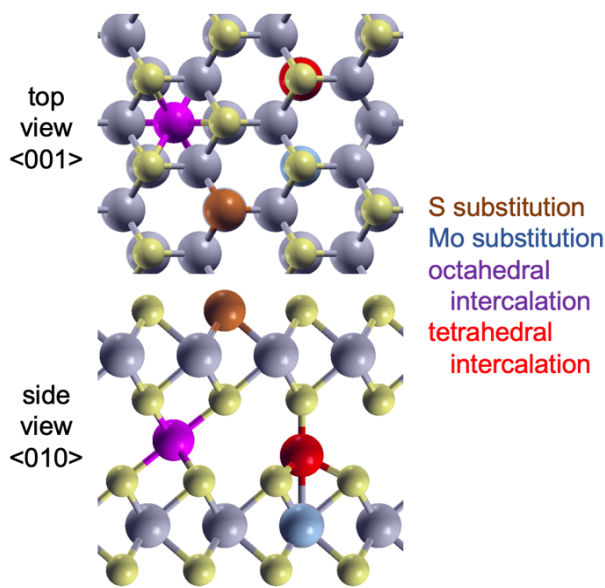

**Figure 2.** Possible doping sites in 2H $MoS_2$: substitution at the S site, substitution at the Mo site, octahedral intercalation, and tetrahedral intercalation. Unsubstituted Mo and S are grey and yellow, respectively.

Re can occur naturally in $MoS_2$. Re substitution of Mo has a low calculated formation energy [27], and it can drive structural changes to 1T or 3R. $ReS_2$ has a similar structure to $MoS_2$ although it does have Re-Re bonds in plane [167]; Re doping by a diffusion process resulted in a 1:100 Re:Mo ratio. No Raman changes were observed, and in particular no sign of $ReS_2$ formation. TEM showed Re in Mo sites, and a small increase in the in-plane lattice spacing, while scanning tunneling microscopy (STM) suggested Mo substitution within grains and intercalation at grain boundaries [177]. Another STM study also showed Mo substitution by Re [178]. Pt doping has been studied for catalysis. Extended X-ray absorption fine structure (EXAFS) showed Pt-S bonds, indicated Mo substitution, and showed no signs of Pt-Pt bonds, indicating that Pt clusters were not present [170].

Nb doping has been extensively studied since it causes little change in the bonding or electronic structure, which is ideal for electronic doping [169,173]. DFT calculations showed alloying is possible with Nb substituting for Mo up to 25% [166]. Nb has the same oxidation state and a similar ionic radius to Mo, and the $NbS_2$ phase exists, albeit with a different stacking. DFT shows the octahedral intercalated site is favored over tetrahedral, but Mo substitution is much more favored. There is a local distortion toward the $NbS_2$ bond lengths [166]. Pairing of Nb dopants in the monolayer is favored by 0.16 eV[169] but only by 0.01 eV in bulk [166]. TEM showed the lattice spacing is preserved and the Raman spectrum showed only subtle changes. Energy-filtered electron transmission microscopy spectrum imaging showed well dispersed Nb, and EXAFS showed the same atomic environment for Nb and Mo [179]. Interestingly, Nb doping can also induce a transformation from the 2H structure to 3R [173].

Ni doping has been investigated for tribology and catalysis. Sputtering can incorporate 5-7% concentrations, and it increases the size of acicular (needle-shaped) crystallites [80]. Low concentrations of Ni increase $MoS_2$ crystal size in hydrothermal synthesis, and change crystallographic orientations. X-ray photoemission spectroscopy showed the presence of $Ni^0$ and $Ni^{2+}$, consistent with intercalation and substitution. High concentrations of Ni cause formation of Ni sulfides instead; $Ni_3S_4$ was seen in XRD and Raman [174], which is the most stable $Ni_xS_y$ phase according to the Materials Project's calculated phase diagram [22,180]. Another XPS study suggested Ni in nanoclusters is neutral but easily oxidizes in air to $Ni^{2+}$ [28]. Mo substitution is suggested by an XRD/AFM study of 10% doping showing the same structure as undoped material [181]. Co doping is also common for catalysis. Co and Ni in $MoS_2$ nanoclusters occur as Mo substituents near the edges, changing them from triangular to truncated shapes, which is seen by STM and predicted by DFT. EXAFS indicates Co and Ni have approximately 5 bonds to S neighbors, consistent with Mo sites near an edge (or a distorted geometry) [182]. Evidence from EXAFS, STEM, and DFT for monolayer $MoS_2$ has shown Co not only in Mo sites but also S sites, and as adatoms in various positions [165].

Doping with Ti and Zr has been found to distort the structure and reduce crystallinity at high doping levels, related to the octahedral coordination in $TiS_2$ and $ZrS_2$. The out-of-plane lattice parameter was not found to change much in XRD. V intercalates and shows expanded layers in XRD, remaining in the 2H polytype. Cr has 2+ and 3+ oxidation states, unlike 4+ for Mo, and therefore causes disordering [167]. Cr doping enhances the $MoS_2$ longitudinal acoustic band in the Raman spectrum, with an intensity linear with doping level. Energy-dispersive X-ray spectroscopy (EDX) showed dispersal of Cr throughout flakes, and the elastic modulus was found to be increased [183].

Group VI elements can straightforwardly substitute for S. $MoSe_{2(1-x)}S_{2x}$ alloys are reported to be favorable [168]. $MoS_{2(1-x)}O_{2x}$ (based on $MoO_4$) has been observed by EXAFS[175] and found to have decreased crystallinity and an expanded interlayer spacing [184]. $MoS_xO_{3-x}$ (based on $MoO_3$) has been identified through Raman and XPS [185]. Se and Te substitution at 25% causes small systematic changes in lattice parameter. AFM showed the same surface structure for Se, but Te was found to result in surface protrusions [181]. Halogens (F, Cl, Br, I) substituting on the S site have been calculated to have large formation energies, but perhaps substitution could occur by filling S vacancies [169], as suggested by experiments [186]. N doping at the S site has been suggested by XPS, with DFT showing N adopting a "groove" position lower than the S atom plane [187]. Phosphorus is believed to substitute for S, as shown in a laser-doping study of the monolayer, in which sulfur vacancies are created as a first step. The long-term stability of photoluminescence suggested substitution by P rather than surface adsorption as adatoms [188]. H or alkali (Li, K, Cs) atoms can adsorb on $MoS_2$ and donate electrons, preferring a location atop the Mo site [169]. Li can also intercalate, driving 2H to 1T crystal structure transformations, and large changes in the Raman spectrum; this process can be used in Li-ion batteries [172].

*7.2 Tribological Properties of Doped $MoS_2$*

Dopants have been used to improve the tribological properties of $MoS_2$ coatings for several decades [167]. As discussed in Section 7.1, dopant atoms can exist in layered structures through substitution of atoms in the lattice, at interstitial sites between atoms in the lattice, intercalated between layers, or as a separate phase dispersed in the material. Which of these is observed depends on the dopant and its concentration, as well as the deposition process by which it was created. Doped coatings are produced using various deposition methods. Sputter deposition, also called sputtering, was one of the earliest techniques and is still commonly used. However, also applied for doped $MoS_2$ coatings are unbalanced magnetron sputtering, ion beam deposition, and pulsed laser deposition [15]. Dopants that have been frequently studied and improve tribological properties of $MoS_2$ are Ni, Cr, Ti, Au, Zr and $Sb_2O_3$. Doped $MoS_2$ coatings have been shown to exhibit better tribological properties (i.e. friction, wear and life) than undoped $MoS_2$ in vacuum, as well as exhibit superior resistance to oxidation which improves performance and storability in air [15].

Understanding the mechanisms by which dopants improve the tribological and oxidation properties of $MoS_2$ is still a topic of active research. In terms of oxidation resistance, multiple mechanisms have been proposed. First, dopants are thought to affect the growth of $MoS_2$ crystallites which reduces $MoS_2$ phase size and in turn increases density [15]. Higher density can reduce crystalline breakage to improve oxidation resistance [189] as well as by decreasing the number of edge sites available to oxidize [15]. Dopants are also proposed to passivate edge sites so that they are less reactive with available oxygen [190]. Lastly, in cases where dopants exists as a separate phase, it has been proposed that the dopant atoms bond with oxygen to form an oxide [191].

The beneficial effects of dopants on friction, wear and coating life have also been explained via multiple mechanisms. However, the most widely cited mechanism is increased hardness [60,62,167,192-195]. It has been proposed that hardness increases as a result of distortion of the $MoS_2$ crystal structure in doped films [60,62]. Hardness directly benefits wear, which is critical for coating life (or endurance) in many applications, including space missions where coatings must function for many years without replacement. Some papers that report improved hardness also observe that dopants can increase the elastic modulus of the coating [193-195]. However, increased hardness is not the only mechanism available, and it has been shown that some dopants decrease hardness yet improve performance [190]. It has been proposed that increased density (sometimes referred to as reduced porosity) can improve friction and wear [196,197]. Also, for some dopants, the primary mechanism is proposed to be presence of the dopant as a separate nanoscale phase which imparts beneficial properties related to the transfer film [192,198,199]. In practice, the mechanism or mechanisms by which a dopant improves coating performance depends on the dopant, its concentration and, to some degree, conditions during coating deposition and operation. For some of the mostly commonly studied dopants – Ti, Au and $Sb_2O_3$ – these effects are discussed in more detail below.

Ti is one of the most widely studied $MoS_2$ dopants [60,80,148,190,193,200,201]. It has been proposed that the Ti atoms are either intercalated between $MoS_2$ layers [62] or, since Ti and Mo have similar ionic radii, substituted with Mo based on observation of a lack of change in crystal c-axis after doping [202]. Measurements of 16% Ti coatings showed that the films are slightly sub-stoichiometric in sulfur and amorphous [201]. Ti-doped coatings have been reported to have a columnar platelet structure, with higher concentrations of Ti corresponding to increasing density [203]. Ti has been found to increase coating hardness, but only up to a limiting concentration, after which additional Ti has a detrimental effect [62,203]. Tribological properties also improve with the amount of Ti in the coating, up to a limiting value that has been reported to be 18% [62]. Ti improves oxidation resistance, as verified by measuring the same friction with 10.8% Ti across a range of humidities [203]. Lastly, it was found that Ti doped $MoS_2$ performance depends on the counterface material, with $ZrO_2$ being best among brass, GCr15 steel, WC and $ZrO_2$ [195].

Another widely studied $MoS_2$ dopant is Au [190,193,196,198,199,204]. Unlike Ti, Au-doped $MoS_2$ consists of domains of nanocrystalline Au particles dispersed in the $MoS_2$ matrix, where the size of the particles depends on temperature [198,199]. Au dopants have been shown to decrease hardness [190], so the mechanism of tribological improvement is different from that of Ti. Instead, it has been proposed that subsurface coarsening of Au nanoparticles provides load support to allow shear of surface $MoS_2$ basal planes [198]. Direct comparison of $MoS_2$, Au and Au-doped $MoS_2$ (at relatively high concentrations >42%) showed the doped coating was superior to both $MoS_2$ and Au alone. Further, it was reported that the optimal amount of Au was dependent on contact stress, where less Au was better at high stress and more Au was better at low stress [199,204]. To explain this observation, it was hypothesized that, at high stress, Au provided the optimum amount of $MoS_2$ in the contact, while at low stress more Au limited the amount/size of $MoS_2$ transferred, providing a thinner, more uniform transfer film [204]. This hypothesis was consistent with the observation that sliding on Au-doped $MoS_2$ results in the formation of pure, crystalline, aligned $MoS_2$ of about 1 nm thickness on the coating [198,199].

Other metallic dopants have been shown to improve coating performance, although they are the focus of fewer studies than Ti and Au. For example, Cr has been shown to improve friction and wear [60,80,190], and the effect was explained by an increase in hardness. The optimum concentration was found to be 16.6% for hardness and 10% for friction and wear [200]. The hardness mechanism was used to explain coating improvements from Zr dopants as well [60,192]. Zr-doped coatings were found to consist of nanocrystalline Zr in an amorphous matrix and optimum performance was observed at a concentration of 10% [192]. Another metal dopant, Al, was found to be similar to Au in terms of mechanisms. Al did not increase hardness, but still improved friction and wear in humid environments. The results were explained as Al gradually becoming alumina, creating a lubricating mix of wear debris in humid conditions [191]. However, another study of Al showed no performance improvement [80]. Lastly, Ni has been shown to improve tribological performance [80,148,189] and particularly to mitigate the increase in friction with decreasing temperature that is commonly observed with both $MoS_2$ and other doped $MoS_2$ coatings [148]. Ni-dopants are currently used in several space components [15] because they improve tribological performance at low temperature [148] and are available at more reasonable cost than some of the other metallic dopant options [80]. In general, metal dopants have been shown to increase performance up to some maximum concentration (e.g. 11% [189]), above which the solubility limit is reached, resulting in layering and poor tribological performance [60].

The last frequently studied dopant is $Sb_2O_3$ [148,189] and, more often, $Sb_2O_3$/Au [148,189,194,205]. Doping with $Sb_2O_3$ results in a fine-grained microstructure and a two-phase system with $MoS_2$ crystallites dispersed in amorphous antimony oxide [189]. The $Sb_2O_3$-doped coatings were found to have increased density and hardness [197] with an optimum concentration of 35% [189]. However, even better tribological performance (particularly wear) has been measured with $Sb_2O_3$ combined with Au to dope the $MoS_2$. For example, coatings with good tribological performance consisted of 11% $Sb_2O_3$ and 7% Au [205]. $Sb_2O_3$/Au-doped $MoS_2$ was found to have an order of magnitude lower wear rate than $Sb_2O_3$-doped $MoS_2$ at room temperature [148]. This feature is particularly important for space applications for which endurance is the most important factor, so $Sb_2O_3$/Au coatings are currently in use for some space components [15].

Several studies have compared multiple dopants. In measurements of friction and life of Ni, Fe, Au, and $Sb_2O_3$ doped coatings, all were better than undoped $MoS_2$ and, among the group, the best was found to be $Sb_2O_3$/Au [189]. Another comparative study of friction and endurance that included Al, Pt, Ag, W, Cr, Co, Ni, and Ti showed some dopants did not improve performance (Al, Pt, Ag, W) while others had a positive effect (Cr, Co, Ni Ti). For those that improved performance, the optimal metal content was found to be 5-8% [80]. Another study of metallic dopants showed no difference between the various dopants [60]. In a comparison between Au, Ti, Cr and $WSe_2$, the lowest friction and life were exhibited by $WSe_2$, followed by Au, Ti and Cr [190]. Another study focused on temperature dependence, measured friction and wear from -80 to 180°C for $Sb_2O_3$/Au, $Sb_2O_3$/C, $Sb_2O_3$, Ti and Ni [148]. They found that $Sb_2O_3$/Au had the lowest friction above -25°C, but Ni had the lowest friction at -80°C. $Sb_2O_3$/Au exhibited the lowest wear rate at room temperature. An inverse relationship between wear rate and the change in friction coefficient with temperature was also observed, i.e. high room-temperature wear corresponded to smaller increase in friction with decreasing temperature [148]. Opposite trends in friction and wear were observed in another study as well [190]. One challenge with comparison between doped coatings is that different deposition methods may be used for different dopants, which can affect the properties of the resultant coating. One study isolated this parameter by comparing undoped $MoS_2$, Ti-doped and $Sb_2O_3$/Au-doped coatings deposited the same way [194]. In terms of wear, $MoS_2$ was best at 30°C and worst at 100°C, $Sb_2O_3$/Au was highly abrasive at 30° but performed well at 100°C, and Ti exhibited good tribological behavior at both temperatures [194].

In summary, dopants have been shown to consistently improve the tribological performance of $MoS_2$ coatings. There are many different types of dopants, but they can be generally categorized into hard materials (Ti, Ni, $Sb_2O_3$) and soft materials (Au, Al). In terms of oxidation resistance, the possible roles of these dopants are to densify the film (limiting the number of reactive sites available), passivate edge sites, and sacrificially bond with oxygen. It is possible that all these mechanisms contribute to improved oxidation resistance of doped $MoS_2$ films. Friction, wear and life of the coatings can also be improved through different mechanisms. However, these mechanisms are highly dependent on the dopant and can be mutually exclusive. Hard materials densify and harden the film, providing direct wear resistance. Soft materials appear to enable formation of thinner and more even $MoS_2$ films on the surface of the coating, indirectly leading to lower wear. Coatings that contain both hard and soft materials, e.g. $Sb_2O_3$/Au, may take advantage of both mechanisms. However, a key observation is that wear is often inversely related to friction, particularly friction at low temperatures. Therefore, dopants should be selected based on the priorities of a specific application.

The research performed on doped $MoS_2$ coatings over the past few decades has clearly demonstrated their great promise. However, as with all new technologies, there are challenges and opportunities. First, scalability is an issue for some deposition techniques [167], which can limit the potential applications for which doped $MoS_2$ may be considered. Second, not all elements can be used for doping, since doping layered structures with elements that exhibit competing bonding coordination can lead to formation of 3D as opposed to layered structures [167]. Third, and probably most significant, massive amounts of testing will be required before sufficient data are available for doped MoS$_2$ coatings to be widely used in engineering applications. One limitation to broad testing is that head-to-head comparison of tribological properties is challenging because different coatings are often produced using different deposition techniques and under inconsistent conditions [194]. Testing for a broad range of applications is also hindered by the fact that many of those applications will operate in extreme environments for long durations (e.g. decade-long space missions in vacuum and extreme temperatures); these conditions are difficult to reproduce in a lab setting. Another challenge is lack of availability of doped MoS$_2$ coatings with controlled composition and deposition. Only a few research labs have the equipment and experience to produce such materials and, while some coatings (Ni-doped and Sb$_2$O$_3$/Au-doped) are commercially available, details about deposition conditions or composition are not usually openly shared. Lastly, given the limitations of experimental testing and the fact that the mechanisms by which these coatings perform their function is hidden from view (i.e. "buried") in sliding interfaces, simulations would be an ideal tool for exploration. However, such simulations require empirical potentials to describe the interatomic interactions and, to model wear, require reactive potentials (which capture bond formation and breaking). At this point, there is no reactive potential available for MoS$_2$ with commonly used dopants. The development of such a model would enable direct interrogation of the doped MoS$_2$ sliding mechanisms, which could provide the knowledge base necessary to tune dopants in order to provide optimized tribological properties for specific applications.

## 8. Conclusions and Outlook

This review provided a focused overview of solid lubrication with MoS$_2$, starting from the fundamentals of its structure and synthesis, followed by an analysis of the conditions under which its use as a solid lubricant is advantageous, and corresponding tribological applications. The current understanding of the fundamental mechanisms underlying the low friction and wear properties of MoS$_2$ was discussed in detail, as well as the dependence of these properties on the operating environment and temperature. Finally, the structure and tribological properties of doped MoS$_2$ were reviewed. The main conclusions of the review and thoughts on emerging research directions are provided below.

There is now agreement regarding the fundamental mechanisms responsible for the lubricative character of MoS$_2$, namely the ability of its basal planes to orient themselves parallel to the sliding direction during sliding and the low shear strength of the interface between basal planes, facilitating inter-plane sliding with minimal frictional resistance to motion. Despite this general agreement, the conclusions are based mostly on investigating MoS$_2$ surfaces *after* the sliding experiments, which does not provide a direct means of studying the details of the dynamic behavior of MoS$_2$ during sliding. As such, great strides in our understanding of MoS$_2$ solid lubrication can be made by devising new experimental methods that provide an *in situ*, high-resolution view of the sliding interface in an MoS$_2$-lubricated contact. While this would be certainly a long-term goal, such an approach would also provide concrete answers to questions about how wear occurs in MoS$_2$. Recent cross-sectional TEM work performed on isolated MoS$_2$ flakes is promising in this regard and future work could build on this achievement to focus on MoS$_2$ coatings that are more application-oriented.

Research shows that the tribological behavior of MoS$_2$ is highly sensitive to the operating environment. In particular, the presence of water or oxygen degrades the excellent lubricative properties of MoS$_2$ observed in vacuum environments. While some studies show that water adsorption plays an important role in the deterioration of the tribological properties of MoS$_2$ around room temperature, at higher temperatures oxidation was shown to be responsible for diminishing tribological performance. Despite agreement regarding the effect of the operating environment on the tribological properties of MoS$_2$, there are still significant discrepancies in the proposed mechanisms, mostly due to difficulties associated with studying different environmental factors (e.g. the presence of water, oxygen or other species) in an isolated fashion. Systematic experiments conducted in well-controlled environments, as well as detailed atomistic simulations employing realistic interaction potentials, can thus provide insight into the environmental dependence of the tribological properties of MoS$_2$, which can eventually enable it to operate reliably in varying environmental conditions.

Based on growing interest in the use of MoS$_2$ in aerospace/space applications where components are expected to operate in a wide range of temperatures (from the cryogenic to several hundred degrees Celsius), the temperature dependence of the tribological properties of MoS$_2$ has been the subject of many studies. Generally, in the absence of wear, friction has been found to decrease with increasing temperature as predicted for a thermally activated process. Wear rate, however, has been found to increase as the temperature increases. Despite many studies reporting observations of the tribological behavior of MoS$_2$ at various temperature ranges, a systematic study of the mechanisms underlying these observations is still lacking. The design and construction of new experimental setups that would allow extended tests of lubricants at extreme temperatures and under controllable environments could prove useful in this regard, especially when they are coupled with atomic-scale simulations that could reveal new insight into the structural and chemical changes that occur at the *buried* sliding interface as a function of temperature.

The improved tribological performance exhibited by doped MoS$_2$ coatings when compared with their pristine counterparts (in particular, under challenging environmental conditions) has enabled their use in critical technologies, including, but not limited to those related to space exploration. Despite this fact, significant limitations/challenges remain related to scalability, the range of materials that can be successfully used for doping, and the lack of availability of relevant testing setups that can successfully simulate, for instance, decade-long missions under vacuum and at cryogenic temperatures. Moreover, while the specific method of deposition and related parameters are known to affect the tribological performance of doped MoS$_2$ coatings significantly, systematic studies evaluating these relations do not currently exist, in part due to the fact that details of composition and deposition of commercially successful coatings are not openly shared. Given the experimental difficulties, simulations arise as a viable route for (i) evaluating and understanding the properties of existing doped MoS$_2$ coatings, especially by studying the buried sliding interface not accessible to conventional experiments, and (ii) predicting new dopants and doping strategies for improved performance. Naturally, the success of such simulations will crucially depend on the availability of accurate, reactive interaction potentials, the development of which could constitute an important direction of research in the near future.

Progress in doped MoS$_2$ tribology will also benefit from advancement in experimental structural characterization. Standard methods for the analysis of periodic crystals, with diffraction of X-rays, electrons, or neutrons, can provide only limited and average information for heterogeneous structures. Most doping studies have not conclusively identified the structure at the dopants. More work employing methods that can provide localized information (such as EXAFS and XPS, and high-resolution STM, AFM, and TEM), in conjunction with DFT calculations, is needed to elucidate dopant locations, local environments, and structural distortions. Study of vibrations by Raman and infrared spectroscopy or neutron scattering, supplemented by theoretical studies, are also a promising way to investigate signatures of dopant concentrations and environments. Correlation of detailed structural characterization with tribology studies is needed to identify the structure-property relations. Knowledge from the extensive literature on MoS$_2$ doping for electronic and catalytic applications can also be leveraged to provide new insight and ideas for tribology. Experimental and theoretical work is then needed to better understand how dopants affect growth processes and morphology, and how the resulting structures can be controlled, to target the structures found to be beneficial to tribological performance.

The rapidly widening range of conditions in which mechanical systems and components have to operate efficiency will certainly maintain and accelerate interest in solid lubrication in the near future. Within this context, the goal of this review was to provide an overview of the properties, applications and mechanisms of MoS$_2$ as a solid lubricant, with the aim of providing a snapshot of our current understanding of this material, as well as inspire future research directions. Focused experimental and computational work are expected to accelerate developments in this field, potentially providing new opportunities for solid lubrication using $MoS_2$ in applications that traditionally employ oils and greases.

**Funding:** This research was supported by the Merced Nanomaterials Center for Energy and Sensing (MACES) via the National Aeronautics and Space Administration (NASA) Grant No. NNX15AQ01. A.M. and M.R.V acknowledge support from the National Science Foundation, award # CMMI-1762384.

**Conflicts of Interest:** The authors declare no conflict of interest.